\documentclass[a4paper]{article}
\usepackage{INTERSPEECH2022}
\usepackage{pifont}
\usepackage{multirow}
\usepackage{xcolor}

\title{A Step Towards Preserving Speakers’ Identity \\ While Detecting Depression Via Speaker Disentanglement}
\name{Vijay Ravi$^\dagger$, Jinhan Wang$^\dagger$, Jonathan Flint$^\S$, Abeer Alwan$^\dagger$ \thanks{This work was funded in part by NIH award number R01MH122569.}}
\address{
  $^\dagger$Dept . of Electrical and Computer Engineering, University of California, Los Angeles, USA\\
  $^\S$Dept. of Psychiatry and Biobehavioral Sciences, University of California, Los Angeles, USA}
\email{(vijaysumaravi@g, wang7875@g, jflint@mednet, alwan@ee).ucla.edu}

\begin{document}

\maketitle
\begin{abstract}
Preserving a patient’s identity is a challenge for automatic, speech-based diagnosis of mental health disorders. In this paper, we address this issue by proposing adversarial disentanglement of depression characteristics and speaker identity.  The model used for depression classification is trained in a speaker-identity-invariant manner by minimizing depression prediction loss and maximizing speaker prediction loss during training. The effectiveness of the proposed method is demonstrated on two datasets - DAIC-WOZ (English) and CONVERGE (Mandarin), with three feature sets (Mel-spectrograms, raw-audio signals, and the last-hidden-state of Wav2vec2.0), using a modified DepAudioNet model. With adversarial training, depression classification improves for every feature when compared to the baseline. Wav2vec2.0 features with adversarial learning resulted in the best performance (F1-score of 69.2\% for DAIC-WOZ and 91.5\% for CONVERGE). Analysis of the class-separability measure (J-ratio) of the hidden states of the DepAudioNet model shows that when adversarial learning is applied, the backend model loses some speaker-discriminability while it improves depression-discriminability. These results indicate that there are some components of speaker identity that may not be useful for depression detection and minimizing their effects provides a more accurate diagnosis of the underlying disorder and can safeguard a speaker’s identity.
\end{abstract}


\section{Introduction}
\label{sec:intro}

Mental health disorders (e.g. Major depressive disorder or MDD) are identifiable  by verbal cues such as monotonic-speech, choice of vocabulary, abnormal discfluencies, etc~\cite{nilsonne1988speech, andreasen1976linguistic}. Previous studies have identified discernible differences between speech patterns of depressed and healthy subjects~\cite{cummins2015review,france2000acoustical}. In addition, collecting speech data has become increasingly easier with the advent of digital voice assistants. As a result, speech-based automatic detection of MDD has received special attention in recent years~\cite{alghowinem2013detecting,ringeval2019avec,low2020automated}. The effectiveness of speech features - spectral~\cite{sanchez2011using,dubagunta2019learning}, prosodic~\cite{yang2012detecting}, articulatory~\cite{seneviratne2022multimodal} and voice-quality~\cite{afshan2018effectiveness} - for classification of MDD, have been analyzed in the past. Lately, deep-learning methods for MDD detection have become popular as they outperform traditional pattern recognition techniques~\cite{ma2016depaudionet, rejaibi2019mfcc, shen2022automatic,chlasta2019automated,harati2021speech}. 

Among others, acoustic features such as  x-vectors~\cite{ravi2022fraug}, i-vectors~\cite{di2021using} and other speaker embeddings~\cite{dumpala2021significance} have been shown to be effective in the diagnosis of a speaker's mental state. These features, however, also carry information about a speaker's identity~\cite{snyder2018x} which can be counter-productive to privacy preservation- a key factor in the adoption of digital mental-health screening systems~\cite{lustgarten2020digital}.  Subsequently, an important question that has remained unanswered in the speech-research community is whether depression detection can be done in a speaker-identity-invariant manner. Additionally, it is still not known if there are components of speech characterizing a speaker that may not be relevant to their mental health status. More recently, two studies introduced algorithms to preserve privacy during depression detection;~\cite{dumpala21_smm} proposed sine-wave speech representation and~\cite{suhasprivacy} used federated learning. However, the performance of depression detection in both studies degrades while preserving patient's privacy. In this paper, the paradigm of adversarial learning to disentangle speaker and depression characteristics is investigated.  


%

In the past, adversarial speaker normalization has been evaluated in the domain of emotion recognition~\cite{yin2020speaker,li2020speaker,gat2022speaker}. In~\cite{yin2020speaker}, the authors perform speaker-invariant domain adaptation on multi-modal features (speech, text, and video) for emotion recognition. In~\cite{li2020speaker}, gradient reversal technique with an entropy loss is proposed to disentangle emotion and speaker information. In~\cite{gat2022speaker}, the authors fine-tune a pre-trained Hubert-base model (\cite{hsu2021hubert}, 300M parameters) with gradient-based adversarial learning.  Fine-tuning such models can require large amounts of in-domain data and be computationally intensive. Moreover, these papers utilize IEMOCAP and MSP-Improv datasets which are mono-lingual and consists of acted audio data~\cite{busso2008iemocap,busso2016msp}. 


In contrast, in this paper, we propose adversarial disentanglement of speaker-identity and depression information, using speech-features only, on datasets with continuous and non-acted speech. In addition to Mel-spectrograms and raw-audio signals, we propose the use of the last-hidden-state of a pre-trained Wav2Vec2.0 model~\cite{baevski2020wav2vec} as the input feature. Wav2vec2.0 models are trained on large amounts of unlabelled data making them robust in various speech processing tasks. The Wav2vec2.0 model is used as a feature extractor, thereby reducing fine-tuning computations. 

Further, unlike prior studies in emotion recognition, we show that the benefits of the proposed method extend to another language (Mandarin). Lastly, we analyze the class separability power of the hidden states of the backend model. We find that while speaker-separability is reduced with adversarial learning, depression-discrimination capability improved. 

To the best of our knowledge, this is the first work to address privacy in depression diagnosis from a speaker-disentanglement perspective. We hypothesize that not all speaker information is useful for depression detection and successfully demonstrate that normalizing irrelevant speaker-related information can result in a more accurate diagnoses, not to mention help protect patients' identity.  



\section{Adversarial Learning}
\label{sec:method}
We propose a loss-based adversarial learning mechanism for speaker-disentangled depression detection. Inspired from the domain-adversarial training proposed in~\cite{ganin2016domain}, our approach involves a loss minimization-maximization technique.

\begin{figure}[ht]
    \centering
    \includegraphics[width=0.85\linewidth]{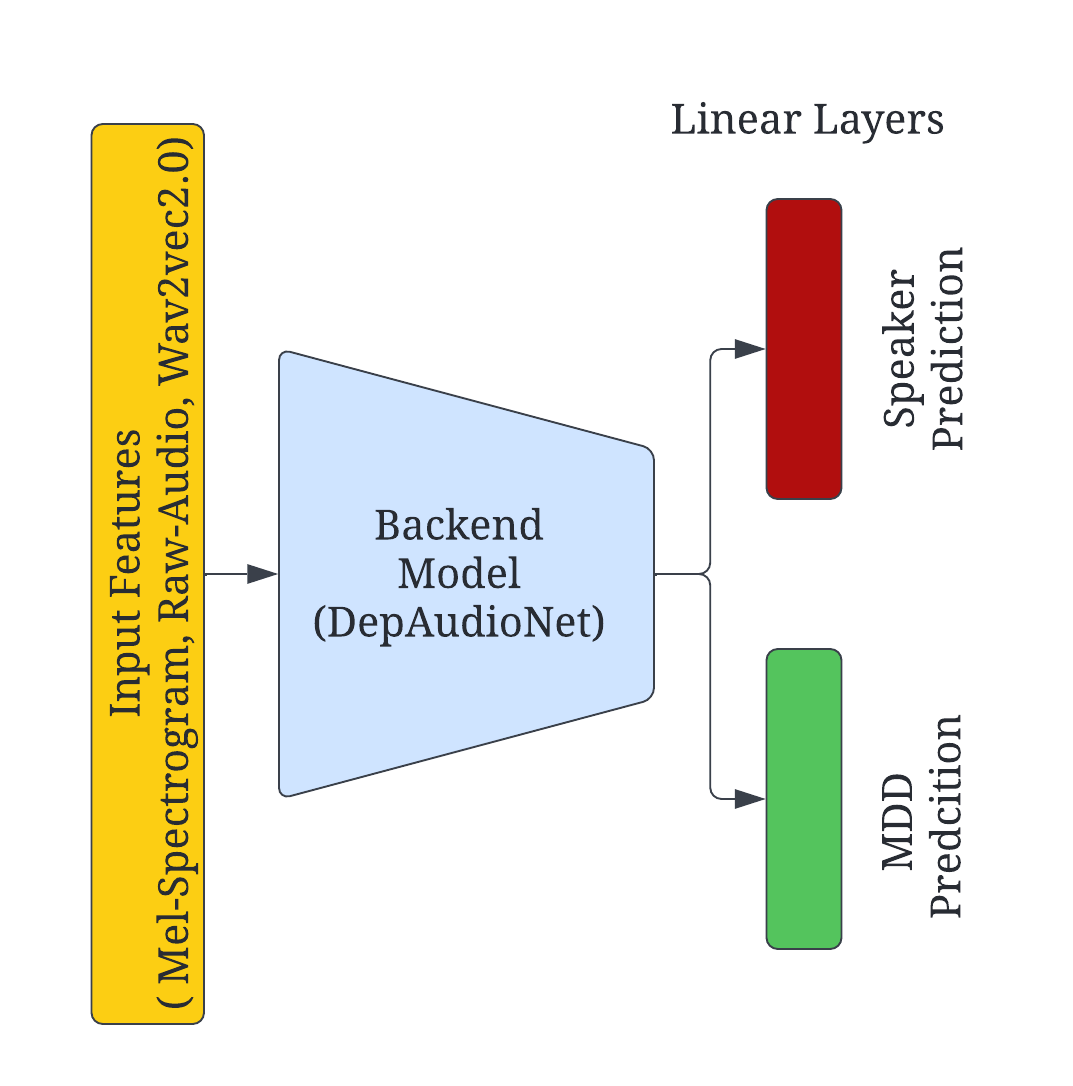}
    \caption{\label{fig:model_arch} Block diagram representing adversarial disentanglement of speaker and depression characteristics.}
\end{figure}

Let the number of unique speakers in the training data be $N$. The loss used for the prediction of MDD binary labels is: 

\begin{equation}
\centering
    L_{MDD}  = -\frac{1}{N} \sum_{i=1}^{N} [Y_i \log{(p_i)} + (1-Y_i) * \log{(1-p_i)}] 
\end{equation}

$Y_i\in({0,1})$ is the class label for the $i^{th}$ speaker and $p_i$ is the probability that speaker $i$ is depressed. The loss for speaker ID prediction is defined as - 

\begin{equation}
\centering
    L_{SPK} = -\frac{1}{N} \sum_{n=1}^{N} [\log{\frac{\exp{(x_{n,n})}}{\sum_{c=1}^{N} \exp{(x_{n,c})}}}],
\end{equation}
\noindent
where  $x_{n,n}$ is the output score that the $n^{th}$ speaker is predicted correctly and $x_{n,c}$  is the score of the $n^{th}$ speaker being predicted as another speaker $c$. 

To train the model in a speaker-identity-invariant manner, during optimization, we minimize the depression loss and maximize the speaker prediction loss.  This can be written as: 

\begin{equation}
\centering
    L_{total} = L_{MDD} - \lambda ( L_{SPK}),
\end{equation}
\noindent
where $\lambda$ is an empirically determined hyperparameter that controls how much of the speaker loss contributes to the total loss.  Initial lambda values were selected to be similar to those reported in the literature (1e-3)~\cite{yin2020speaker}. We experimented with higher and lower values and chose the best performing lambda values. By this process, we force the model to focus more on  depression-discriminatory information and ignore some speaker-discriminatory information, thereby making the model invariant to changes in some speaker-specific characteristics. 


\section{Experimental Details}
\label{sec:exp_details}

Experiments were conducted using the DAIC-WOZ (English,~\cite{valstar2016avec}) and CONVERGE (Mandarin,~\cite{li2012patterns}) datasets. The backend model was a modified version of the DepAudioNet~\cite{ma2016depaudionet} which was trained using Mel-spectrograms, raw-audio signals and the last-hidden-state of Wav2vec2.0 as input features. Details of the datasets, model parameters, input features, and evaluation metrics used are described in this section.

\subsection{Datasets}
\subsubsection{DAIC-WOZ}
The Distress analysis interview corpus wizard of Oz (DAIC-WOZ)~\cite{valstar2016avec} database comprises audio-visual interviews of 189 participants, male and female, who underwent evaluation of psychological distress. Each participant was assigned a self-assessed depression score through the patient health questionnaire (PHQ-8) method~\cite{kroenke2009phq}. Audio data belonging only to the participants were extracted using the time-labels provided with the dataset. Recordings from session numbers 318, 321, 341 and 362 were excluded from training because of time-labelling errors. This dataset consists of 58 hours of audio data, sampled at 16kHz. Data partitioning is the same as that provided with the database description (107 speakers for training and 35 speaker for evaluation).

\subsubsection{CONVERGE}
The CONVERGE (China, Oxford and Virginia Commonwealth University Experimental Research on Genetic Epidemiology) dataset~\cite{li2012patterns} comprises of audio interviews of 7959 female participants being interviewed by a trained interviewer. This study focused on subjects with increased genetic risk for MDD, and to obtain a more genetically homogeneous sample, only women participants were recruited. The diagnoses of depressive disorders were made with the Composite International Diagnostic Interview (Chinese version)~\cite{TerSmitten1998}. Audio recordings are sampled at a rate of 16kHz. In this study, a subset of this dataset is used with 1185 randomly chosen speakers in the training partition and the remaining 507 speakers in the evaluation partition. This database is characterized by a large degree of phonetic and content variability.  

\subsection{Model - DepAudioNet}
DepAudioNet was chosen as a baseline because its code is publicly available in addition to its high accuracy of depression classification. Model implementation was based on~\cite{bailey2021gender} and network parameters such as the number of hidden layers, learning rate,  and dropout probability etc. were chosen empirically. Kernel size and stride of the \textit{Conv1D} layer are denoted by $K$ and $S$, respectively. Hidden state dimension of the recurrent LSTM layers is denoted by $H$.

For the DAIC-WOZ dataset, the model with Mel-spectrograms as input consisted of one \textit{Conv1D} layer ($K=3$, $S=1$) and four unidirectional LSTM layers ($H=128$). In case of raw-audio signals, two \textit{Conv1D} layers ($K_1=1024$, $S_1=512$, $K_2=3$, $S_2=1$) and two LSTM layers ($H=128$) were used. For Wav2vec2.0 features, the model consisted of an input LSTM layer ($H=256$) followed by five hidden LSTM layers with the same $H$ as the input layer. For the CONVERGE data and Mel-spectrograms, the model consisted of two \textit{Conv1D} layers ($K_1=3$, $S_1=1$, $K_2=3$, $S_2=1$) and four unidirectional LSTM layers ($H=128$). In case of CONVERGE and raw-audio signals,two \textit{Conv1D} layers ($K_1=1024$, $S_1=512$, $K_2=3$, $S_2=1$) and four LSTM layers ($H=512$) were used. For Wav2vec2.0 features with CONVERGE data,  the model consisted of an input LSTM layer ($H=256$) followed by five hidden LSTM layers with the same $H$ as the input layer.

\textit{Conv1D} layers were followed by ReLU non-linearity, a dropout layer and a max-pooling layer with a kernel of size 3. For every model configuration, the final layers were fully connected layers to generate the predictions for MDD and speaker labels. Based on the number of speakers in the training set, output dimensions for speaker labels were 107 for experiments with DAIC-WOZ and 1185 for CONVERGE. For MDD prediction, a sigmoid activation was used with binary cross entropy loss, while for speaker ID prediction, cross entropy loss was used without any output activation.  

\subsection{Input Features}

Three feature sets were analyzed -- 1) 40-dimensional Mel-spectrograms extracted using a Hanning window of length w = 1024 samples (64ms), and hop size h = 512 samples (32ms), 2) raw-audio signal, and 3) the last-hidden-state of Wav2vec2.0 model~\cite{baevski2020wav2vec}. For the DAIC-WOZ data, Wav2vec2.0 base model was used where the hidden state dimension was 768. For CONVERGE, Wav2vec2.0 large model (XLSR-Mandarin) was used where the hidden state dimension was 1024~\cite{conneau2020unsupervised}. Mean-variance normalization was applied to Mel-spec and raw-audio features. Wav2vec2.0-base and XLSR-Mandarin models from the Hugging Face library~\cite{patrickvon2021wav2vec2, grosman2021wav2vec2} were used. 

In the case of the DAIC-WOZ dataset, training data were pre-processed by random cropping and sampling~\cite{ma2016depaudionet}. Each utterance was randomly cropped to fragments (equal to the length of the shortest utterance) and every fragment was segmented into multiple segments. Segment lengths were  3.84s each which translates into 120 frames for Mel-spectrogram, 61440 samples for raw-audio and 200 frames for Wav2vec2.0 features. A training subset was generated by randomly sampling,  without replacement, an equal number of depression and non-depression segments. In each experiment, five separate models were trained using a randomly generated training subset and the final prediction was averaged across those five models. On the contrary, for the CONVERGE dataset, segments were generated without random cropping and sampling with the segment length the same as before, and each experiment was performed by training one model using all of the training data.

\subsection{Evaluation Metrics}

Depression classification was evaluated using the F1-score to account for both false positives and false negatives~\cite{chinchor1992muc}.

The J-ratio is a metric used to measure class separability of a given set of features~\cite{fukunaga2013introduction} and is computed as follows:

\begin{equation}
    S_W = \frac{1}{N} \sum_{i=1}^{N} R_i,
\end{equation}

\begin{equation}
    S_B = \frac{1}{n}\sum_{i=1}^{N} (M_i - M_o)(M_i-M_o)^T,
\end{equation}

\begin{equation}
    J = trace[(S_W+S_B)^{-1}S_B],
\end{equation}

\noindent
where $N$ is the total number of speakers, $R_i$ is the covariance matrix for the $i^{th}$ speaker, $M_i$ is the mean vector for the $i^{th}$ speaker, and $M_o$ is the mean of all $M_i$s. $S_W$ and  $S_B$ are the within- and between-class scatter matrices, respectively. Hidden states of all recurrent layers (LSTMs) of the DepAudioNet model were extracted and used as inputs in the calculation of J-ratios. This metric was used to examine speaker separability~\cite{guo2016speaker}.

\section{Results and Discussion}
\label{sec:results}
Results are presented and discussed in two steps. First, performance of the proposed method relative to the baseline system is presented for each of the three input features and the two datasets. Then, through speaker-separability analysis, it is shown that the proposed method indeed normalizes some speaker information resulting in better MDD classification. Relative improvements, unless specified, are statistically significant~\cite{mcnemar_note_1947}. 

\subsection{Adversarial Learning}

Table~\ref{tab:adversarial_daic} shows the performance of the proposed approach using the DAIC-WOZ dataset in terms of the F1-score for the Non-Depressed Class (ND), the Depressed class (D) and the average F1-score (F1-Avg) across both classes.  In the case of Mel-spectrograms, the baseline DepAudioNet model~\cite{ma2016depaudionet} has an F1-Avg of 0.619 which is in agreement with previous work~\cite{bailey2021gender}. When adversarial speaker-invariant training is applied, the overall relative performance improvement is 4.36\% where F1-ND increased by 3.6\% and F1-D by 5.06\%. The baseline performance with raw audio signals as input is better than the Mel-spec baseline. In this case, with adversarial learning, while F1-ND is reduced by 6.8\%, F1-D increased by 16\% resulting in an overall relative improvement of  2.16\%. With Wav2vec2.0 features, in-spite of a very high baseline performance (F1-Avg = 0.686), the proposed method provides overall improvements of 0.87\% resulting in the best F1-Avg score of 0.692. 

\begin{table}[ht]
\centering
\caption{F1-scores for DepAudioNet using the DAIC-WOZ dataset for three features, with and without adversarial speaker disentanglement. F1-Avg is the average of F1-scores for non-depressed (F1-ND) and depressed (F1-D) classes. +Adv denotes adversarial training. $\lambda$ values used for disentanglement are mentioned in parenthesis.}
\label{tab:adversarial_daic}
\resizebox{\linewidth}{!}{%
\begin{tabular}{ccrrr}
\hline 
\hline
\textbf{Input Feature} & \textbf{+Adv} & \multicolumn{1}{c}{\textbf{F1-Avg}} & \multicolumn{1}{c}{\textbf{F1-ND}} & \multicolumn{1}{c}{\textbf{F1-D}} \\ \hline \hline

Mel-spec & No                      & 0.619          & 0.706 & 0.533 \\
Mel-spec & Yes ($\lambda$ = 5e-6) & 0.646          & 0.732 & 0.560 \\ \hline 
Raw-audio       & No                      & 0.646 & 0.779 & 0.512 \\
Raw-audio       & Yes ($\lambda$ = 1e-6) & 0.660          & 0.726 & 0.594 \\ \hline
Wav2vec2.0      & No                      & 0.686 & 0.804 & 0.567 \\
Wav2vec2.0      & Yes ($\lambda$ = 1e-3) & 0.692          & 0.808 & 0.576 \\ \hline \hline
\end{tabular}%
}
\end{table}

Similar improvements are observed using the CONVERGE dataset. As seen in Table~\ref{tab:adversarial_converge}, for Mel-spectrograms, relative performance increases by 1.25\%, from 0.879 for the baseline to 0.890 for the proposed method. Improvements for raw-audio are higher; baseline system has an F1-Avg of 0.829 whereas the proposed method has an F1-Avg of 0.857 (relative improvement of 3.34\%). Lastly, for the features extracted from the XLSR-Mandarin model,  a relative improvement of 0.33\% (not statistically significant) is observed -  F1-score for baseline is 0.912 and 0.915 for the proposed adversarial disentanglement, respectively. For both datasets, features extracted from Wav2vec2.0 models, which are trained in an unsupervised manner and fine-tuned for speech recognition, performed the best. 

Another important observation to note is that adversarial disentanglement of speaker and depression was sensitive to $\lambda$ values and in most experiments, a very small value resulted in significant improvements.

\begin{table}[ht]
\centering
\caption{F1-scores for DepAudioNet using the CONVERGE dataset for three features, with and without adversarial speaker disentanglement. `+Adv` denotes adversarial training.}
\label{tab:adversarial_converge}
\resizebox{\linewidth}{!}{%
\begin{tabular}{ccrrr}
\hline 
\hline
\textbf{Input Feature} & \textbf{+Adv} & \multicolumn{1}{c}{\textbf{F1-Avg}} & \multicolumn{1}{c}{\textbf{F1-ND}} & \multicolumn{1}{c}{\textbf{F1-D}} \\ \hline \hline

Mel-spec & No                      & 0.879	& 0.890 & 0.868 \\
Mel-spec & Yes ($\lambda$ = 5e-5) & 0.890	& 0.903 &	0.877 \\ \hline 
Raw-audio       & No            & 0.829 &	0.832	& 0.826  \\
Raw-audio       & Yes ($\lambda$ = 2e-4) & 0.857 &	0.870 &	0.844 \\ \hline
XLSR-Mandarin      & No                      & 0.912  & 0.921 & 0.903  \\
XLSR-Mandarin      & Yes ($\lambda$ = 2e-4) &     0.915      &  0.925&  0.906\\ \hline \hline
\end{tabular}%
}
\end{table}

Relative improvements in system performance are summarized in Table~\ref{tab:f1_change}. The proposed method improves depression prediction score for every feature across two different datasets. More importantly, prediction performance improves for both ND and D classes (except for DAIC-WOZ raw-audio), demonstrating that the proposed method can improve precision and recall of automatic MDD diagnosis in most cases. Improvements, however, are marginal for content-rich features such as Wav2vec2.0 base and XLSR-Mandarin. It maybe that since these model are fine-tuned for speech recognition tasks, much of the speaker-information is already lost during feature extraction resulting in smaller improvements. 

\begin{table}[ht]
\centering
\caption{Percentage change (relative) in F1-Avg after adversarial disentanglement for three input features using DAIC-WOZ and CONVERGE datasets. $^*$ indicates that the change is not statistically significant.
}
\label{tab:f1_change}
\begin{tabular}{ccc}
\hline
\hline
\textbf{Dataset} & \textbf{Input Feature} & \multicolumn{1}{c}{\textbf{\begin{tabular}[c]{@{}c@{}}\% Change \\ in F1-Avg\end{tabular}}} \\ \hline \hline
\multirow{3}{*}{DIAC-WOZ}                     & Mel-spec & +4.36   \\ 
                                              & Raw-audio       & +2.16   \\  
                                              & Wav2vec2.0      & +0.87   \\ \hline
\multicolumn{1}{l}{\multirow{3}{*}{CONVERGE}} & Mel-spec & +1.25   \\  
\multicolumn{1}{l}{}                          & Raw-audio       & +3.34          \\  
\multicolumn{1}{l}{}                          & XLSR-Mandarin      & +0.33$^*$        \\ \hline \hline
\end{tabular}%
\end{table}

\subsection{Speaker Separability Analysis}

Speaker separability, measured in terms of J-ratio, along with F1-Avg scores for MDD prediction are presented in Tables~\ref{tab:jratio_daic} and~\ref{tab:jratio_converge} for the DAIC-WOZ and CONVERGE datasets, respectively. A higher J-ratio indicates better speaker discrimination. For the DAIC-WOZ dataset, the proposed method reduces J-ratio for every feature while improving the F1-Score. Interestingly, configurations with lower speaker separability performed better in depression classification - for example baseline systems vs. adversarial disentanglement and Wav2vec2.0 vs. Mel-spectrogram.  

\begin{table}[htbp]
\centering
\caption{J-ratios and F1-Avg scores for three input features with and without adversarial learning on DIAC-WOZ data. `+Adv` denotes adversarial training}
\label{tab:jratio_daic}
\begin{tabular}{ccc}
\hline \hline
\textbf{Input Features} & \textbf{J-Ratio} & \textbf{F1-Avg} \\ \hline \hline
Mel-spec       & 4.81 & 0.619  \\
Mel-spec+Adv   & 4.60 & 0.646  \\ \hline \hline
Raw-audio      & 2.17 & 0.646  \\ 
Raw-audio+Adv  & 1.84 & 0.660 \\ \hline \hline
Wav2vec2.0     & 1.56 & 0.686 \\ 
Wav2vec2.0+Adv & 1.52 & 0.690 \\ \hline \hline
\end{tabular}%
\end{table}

\begin{table}[htbp]
\centering
\caption{J-ratios and F1-Avg scores for three input features with and without adversarial learning on CONVERGE data. `+Adv` denotes adversarial training}
\label{tab:jratio_converge}
\begin{tabular}{ccc}
\hline \hline
\textbf{Input Features} & \textbf{J-Ratio} & \textbf{F1-Avg} \\ \hline \hline
Mel-spec       & 42.66 & 0.879  \\
Mel-spec+Adv   & 40.61 & 0.890 \\ \hline \hline
Raw-audio      & 302.3 & 0.829  \\ 
Raw-audio+Adv  & 74.64 & 0.857  \\ \hline \hline
XLSR-Mandarin     & 75.54 &  0.912\\ 
XLSR-Mandarin+Adv & 71.62 &  0.915 \\ \hline \hline
\end{tabular}%
\end{table}

Contrary to DAIC-WOZ, the J-ratio for XLSR-Mandarin using the CONVERGE data was higher than those for Mel-spec. It is possible that Wav2vec2.0 captures room acoustics resulting in a higher speaker separability. Nevertheless, trends are  similar, in that the J-ratio for speaker separation reduces for every feature set when the proposed method is applied. Comparing the improvements from Table~\ref{tab:f1_change} with J-ratios presented in Tables~\ref{tab:jratio_daic} and~\ref{tab:jratio_converge}, it can be seen that performance gains are maximum for features with the highest baseline speaker J-ratio; Mel-spec in DAIC-WOZ (+4.36\% improvement and baseline J-ration of 4.81) and Raw-audio in CONVERGE (+3.34\% improvement and baseline J-ratio of 302.3). These results confirm our hypothesis that speaker-identity contains some components irrelevant to the mental state of that person and training the model to be invariant to such characteristics can lead to better depression detection.

\section{Conclusion and Future Work}
\label{sec:conclusion}
In the past, features such as x-vectors, i-vectors, etc. have been shown to be useful for depression detection. These features, despite their effectiveness, contain information about the speaker's identity. Excessive dependence on speaker-identity features can harm the privacy factor of an MDD diagnosis system, an important attribute towards the adoption of speech-based assessment methods.  Naturally, a question that arises is whether depression detection can be done in a speaker-invariant manner.

In this paper, we attempt to address this important question by proposing adversarial disentanglement of speaker-identity and depression status. The proposed method illustrates that speaker-identity invariant models can provide better MDD classification performance across multiple features and on multi-lingual datasets. The results presented in this paper support the hypothesis that when correlates of speaker identity, irrelevant to a subject's mental state, are partially normalized, depression diagnosis is more accurate. 

Future work will investigate the specifics of speaker-related information that are or are not useful for depression detection, and will compare the effectiveness of content-rich vs. speaker-information-rich features.  Other metrics for speaker recognition, such as multi-class accuracy, will also be analyzed.

\bibliographystyle{IEEEtran}

\bibliography{mybib}

\end{document}